# Hierarchical emotion-recognition framework based on discriminative brain neural network topology and ensemble co-decision strategy

Cunbo Li#, Peiyang Li#, Yangsong Zhang, Ning Li, Yajing Si, Fali Li, Dezhong Yao and Peng Xu*

*Abstract*—Brain neural networks characterize various information propagation patterns for different emotional states. However, the statistical features based on traditional graph theory may ignore the spacial network difference. To reveal these inherent spatial features and increase the stability of emotional recognition, we proposed a hierarchical framework that can perform the multiple emotion recognitions with the multiple emotion-related spatial network topology patterns (MESNP) by combining a supervised learning with ensemble co-decision strategy. To evaluate the performance of our proposed MESNP approach, we conduct both off-line and simulated on-line experiments with two public datasets i.e., MAHNOB and DEAP. The experiment results demonstrated that MESNP can significantly enhance the classification performance for the multiple emotions. The highest accuracies of off-line experiments for MAHNOB-HCI and DEAP achieved 99.93% (3 classes) and 83.66% (4 classes), respectively. For simulated on-line experiments, we also obtained the best classification accuracies with 100% (3 classes) for MAHNOB and 99.22% (4 classes) for DEAP by proposed MESNP. These results further proved the efficiency of MESNP for structured feature extraction in mult-classification emotional task.

## I. INTRODUCTION

EMOTION processing plays a crucial role in meditation and social behavior of human beings [5]. A series of studies have converged to evidences that emotional intelligence is one of the most important intelligence in humans' daily life rather than logical intelligence [8]. *Scientific American* reported that whether we can detect human emotions automatically is one of the twenty big questions about the future of humanity [10]. In the last decade, researches have proposed a large variety of strategies to automatically detect different emotional states through physiological signals [12]. These studies aim to enhance the capability of affective brain computer interfaces (aBCI) to effectively detect, process and respond to users affective state [9, 13]. In fact, emotional interaction between humans and peripheral equipment occupies particular role in the design and implementation of aBCI systems [14, 15]. In recent decades, many efforts have been made to enhance the aBCI systems with the ability to communicate and feedback to user's affective states naturally [16]. Actually, emotion refers to both physiological and psychological activities and reflects the external and internal activities of human beings [17, 18], which can usually serve as features to identify different emotional states [19, 20].

Although emotional recognition based on multi-mode audio and visual features, such as the face, voice, and gesture, has achieved great success, there still exist many challenges due to the fact that emotions vary across time, context, space, language, culture, and races [21]. For example, the detection based on face or gesture are sensitive to pose, clutter and lighting conditions, and the auditory noise has a great influence when recognition is based on voice [22]. Moreover, those audio and visual information based emotion recognition is prone to disturbance by the subject's intention to disguise. Increasing studies have found that emotion recognition based on the electrophysiological signals, such as electroencephalogram (EEG), functional Magnetic Resonance Imaging (fMRI), electrocardiogram (ECG) and electromyogram (EMG), are more reliable than that based on audio and visual signals, because those physiological signals can reflect the true emotions even if subjects deliberately disguise [23]. Among them, due to such merits as the excellent temporal resolution reflecting the dynamic changes in the brain [24] and the easy setup as well, EEG is potential to serve as the signal source for emotion recognition. Based on the EEG energy distribution difference between different emotional states, various feature extraction algorithms have been proposed such as power spectrum (PS), power spectral density (PSD), differential entropy (DE), differential asymmetry (DASM), and rational asymmetry (RASM) for emotion recognition [2, 12, 25]. Koelstra and Patras fused the PSD with facial expressions features for implicit affective tagging and achieved the

Manuscript submitted. This work was supported by the National Natural Science Foundation of China (#U19A2082, # 61961160705), the National Key Research and Development Plan of China (#2017YFB1002501), the Key R&D Program of Guangdong Province, China (#2018B030339001).

Cunbo Li, Peiyang Li, Yangsong Zhang, Ning Li, Yajing Si, Fali Li, Dezhong Yao and Peng Xu are with the Clinical Hospital of Chengdu Brain Science Institute, MOE Key Lab for Neuroinformation and School of Life Science and Technology, University of Electronic Science and Technology of China, Chengdu, 610054, China. Peiyang Li is now working in the School of Bioinfomatics, Chongqing University of Posts and Telecommunications, Chongqing, 400065, China. Yangsong Zhang is also with the School of Computer Science and Technology, Southwest University of Science and Technology, Mianyang 621010, China. Corresponding author: Peng Xu, email: xupeng@uestc.edu.cn, Tel: 86-028-83206978.

satisfactory experimental result [2]. Zhao et al. studied the possible relationship between emotion and personality inference with EEG, and predicted personality traits in five dimensions with PSD feature, achieving the best 86.11% accuracy in the dimension of agreeableness [26]. Zheng et al. used DE feature to emotion classification and achieved 70.58% accuracy with all 62 channels of EEG data [10]. However, these features mainly reflect the local information of brain activities, ignoring the information propagation patterns between multiple brain regions that is important to form the human emotions.

In essence, emotion processing in the brain relates to a complex dynamic interactive process involving many brain regions [27-30]. Therefore, the more optimal feature extraction methods should be capable of capturing these interaction patterns in brain. In recent years, the functional connections among different brain regions have been proven to hold close relationship with different emotional states [24, 30]. Y.-Y. Lee and S. Hsieh classified different emotional states with the EEG-based functional connectivity patterns and found that the functional connections are significantly different when the emotional states changes [30]. Li et al adopted a multiple feature fusion approach to combine the activation patterns and connection patterns for emotion recognition, and the performed analysis has shown the functional connection patterns can significantly enhance the classification performance [31]. Y. Dasdemir et al analyzed the functional brain connections for positive and negative emotions states and reported that the control of emotion may be related with the functional connectivity of left frontal electrodes [24]. These findings highlight the importance of the brain functional connectivity for the emotion related researches.

However, the statistical measurements of functional networks though determined by the spatial network topology may fail to reflect the intact spatial information of networks [32]. In fact, identifying the essential spatial network topologies may be potentially helpful to differentiate different emotions. In this work, we proposed a hierarchical framework that can perform the multiple emotion recognitions with the multiple emotion-related spatial network topology patterns (MESNP) by combining a supervised learning with ensemble co-decision strategy. To evaluate the validity and stability of the proposed method, we used two public emotion EEG datasets, i.e., MAHNOB-HCI that provided by M. Soleymani et al [33] and DEAP by Koelstra et al [34] to verify the proposed method. For MAHNOB-HCI, there are three emotion labels, namely negative, neutral and positive. For DEAP, there are four emotion labels, which includes low arousal-low valence (LALV), high arousal-low valence (HALV), low arousal-high valence (LAHV), and high arousal-high valence (HAHV), according to the ratings of valence-arousal (VA) space.

The main contributions of this paper can be summarized as two aspects:

1) We proposed a hierarchical framework to realize the multiple emotion recognition by combining a supervised learning with an ensemble co-decision strategy based on the topological patterns of emotions.

2) A simulated on-line emotion detection system was established for real time emotional recognition, which could also be applied to clinical diagnosis and intervention.

The remainder of this paper is as follows. In Section 2, we briefly introduced the mostly adopted features for emotional recognition. In Section 3, the structure of proposed MESNP are presented elaborately. In Section 4, both off-line and on-line experiments on MAHNOB and DEAP are conducted to evaluate the efficiency of our proposed emotional recognition strategy. An elaborate analysis and discussion on experiment results are given in Sections 5 and 6 to demonstrate the robustness of the proposed method. Finally, conclusions and future works are described in Section 7.

## II. RELATED WORKS

Currently, the widely used EEG features for emotional recognition are generally derived from the spectrum power represented by power spectral density (PSD) and differential entropy (DE). In this work, these two widely applied features were utilized as the benchmarks for comparison, and the details of these two methods are introduced as follows.

### A. Power Spectral Density

Spectrum analysis is widely used in diverse fields such as pattern recognition and signal processing [35] which is usually carried out by the estimation of PSD. In this work, the Welch algorithm was adopted to estimate PSD features in theta, alpha, beta and gamma bands, respectively. The power spectral density $Psd(f)$ of EEG signal $X(t)$ is calculated by the periodogram method, shown as follow:

$$Psd(f) = \frac{1}{TW_{in}} \left| \sum_{t=0}^{T-1} X(t) F(t) e^{-j2\pi ft} \right|^2 \quad (1)$$

where $j$ is an imaginary unit, $T$ is the length of time series, and $F(t)$ is the Hamming window function to reduce spectral leakage. Besides, $W_{in}$ is the regularization coefficients of windows defined as follow, which is used to reduce the influence of window function on the power spectrum estimated.

$$W_{in} = \frac{1}{T} \sum_{t=0}^{T-1} F^2(t) \quad (2)$$

### B. Differential Entropy

DE is proposed to extract features from EEG data [36, 37], which is based on the Shannon entropy that defined as

$$h(X) = -\int_X f(x) \log(f(x)) dx \quad (3)$$

where $f(x)$ is the probability density function of $X(t)$. We assume that the EEG signals are the time series of $X(t)$, which obeys the Gaussian distribution of $N(\mu, \delta^2)$, its differential entropy is defined as [9]

$$h(X) = -\int_{-\infty}^{+\infty} \frac{1}{\sqrt{2\pi\delta^2}} exp\frac{(x-\mu)^2}{2\delta^2} \log \frac{1}{\sqrt{2\pi\delta^2}} exp\frac{(x-\mu)^2}{2\delta^2} dx$$
$$= \frac{1}{2}\log 2\pi e\delta^2 \quad (4)$$

where $X(t)$ follows the Gaussian distribution $N(\mu,\delta^2)$, and $\pi$ and $e$ are constants.

III. METHODS

In this work, motivated from the network difference involved in different emotions processing, we proposed a hierarchical emotional recognition framework based on spatial network patterns as shown in Fig.1. This procedure consists of two sub-procedures, i.e., training and prediction. As shown in Fig.1, the original EEG data was divided into training and testing dataset firstly, and then the network analysis method was adopted to construct functional brain neural networks for training and testing sets. For the training procedure, we divided the training set into pair-wise groups according to label information and a supervised learning is adopted to extract the corresponding MESNP filters and training features for each pair-wise group respectively. Finally, the ensemble SVM system was trained from each group MESNP features. During the prediction procedure, each brain neural network sample would be fed into every group MESNP filters to extract the corresponding MESNP features, and these MESNP features were inputted into the corresponding trained SVMs to predict labels. Finally, the final emotion label would be predicted by the result of ensemble co-decision voting of all the SVM outputs. Compared to the existing emotion prediction approaches, the improvement of proposed one can be attributed to the combination of a supervised learning strategy and ensemble co-decision to extract the discriminative spatial network patterns of multi-classes emotions. The overall procedure for proposed MESNP is summarized in Algorithm 1. In the following sub-sections, we will introduce the related aspects in detail, and in order to facilitate reading, the nomenclature mentioned in this section are shown in Table I.

| Algorithm 1 MESNP realization |
|---|
| 1: Input: EEG Dataset $G$ |
| 2: Output: System predicting accuracy |
| 3: Preprocessing and segmenting dataset $G$ |
| 4: Dividing all segments into training and testing datasets |
| 5: Network analysis with training and testing datasets, respectively |
| 6: Training process. Including spatial filters learning, extraction of training **MESNP** features, and ensemble **SVM** system training |
| 7: Testing process. Including extraction of the testing **MESNP** features by **MESNP** spatial filters and ensemble co-decision based prediction |

A. Brain neural network construction

In current work, we estimated functional brain neural network with Phase Locked Value (PLV) [38]. According to the Gabor analysis [39], for the time signal $X(t)$, we could create an analysis signal $H(t)$ by $X(t)$ and it's Hilbert transform signal as [40]

$$H(t) = X(t) + iX_h(t) = A(t)e^{-i\varphi(t)} \quad (5)$$

where $A(t)$ is the instantaneous amplitude of $X(t)$, and $\varphi(t)$ is the instantaneous phase of $X(t)$. The Hilbert transform of $X(t)$ is as follows,

$$X_h(t) = \frac{1}{\pi}P.V.\int_{-\infty}^{\infty}\frac{X(t)}{t-\tau}d\tau \quad (6)$$

where $P.V.$ is the Cauchy principal value[40].

The phase $\varphi(t)$ and amplitude $A(t)$ of signal $X(t)$ can be uniquely determined by (5) and (6). The calculation of the instantaneous phase is as

$$\varphi(t) = \arg[X(t)] = \arctan\frac{X_h(t)}{X(t)} \quad (7)$$

From (5)-(7), we can respectively calculate the instantaneous phases of two different channel EEG time signals $X_{N_1}(t)$, $X_{N_2}(t)$ ($N_1, N_2 \in [1,32]$), and then calculate the Phase Locked Value of the signal $X_{N_1}(t)$ and $X_{N_2}(t)$ according to (8) as

$$Plv = \sqrt{\left(\frac{\sum_{1}^{T}\cos\left(\varphi_{X_{N_1}}(t) - \varphi_{X_{N_2}}(t)\right)}{t}\right)^2 + \left(\frac{\sum_{1}^{T}\sin\left(\varphi_{X_{N_1}}(t) - \varphi_{X_{N_2}}(t)\right)}{t}\right)^2}$$
(8)

After calculating the PLV for each paired EEG channels, we can get the weighted adjacency matrix to reflect the phase couplings among the recorded EEG signals. Due to the fact that both the two public emotion datasets used 32-channel EEG acquisition systems, the PLV brain neural networks are $32\times32$ weighted adjacency matrix, i.e., PLV brain neural network consists of 32 nodes.

B. Network Properties

There are four mostly used statistical brain neural network measurements, i.e., clustering coefficients ($Cc$), the shortest path length ($L$), global efficiency ($Ge$) and local efficiency ($Le$)[41, 42]. Here, Let $c_{N_1N_2}$ be the edge linkage strength between vertices $N_1$ and $N_2$, $d_{N_1N_2}$ be the shortest weighted path length between vertices $N_1$ and $N_2$, $\Psi$ be the set of nodes in brain neural network, and $N$ ($N=32$) be the node number of brain neural network, these four network properties were defined as follows:

$$Cc = \frac{1}{N}\sum_{N_1 \in \Psi}\frac{\sum_{N_2,N_3 \in \Psi}\left(c_{N_1N_2}c_{N_1N_3}c_{N_2N_3}\right)^{1/3}}{\sum_{N_2 \in \Psi}c_{N_1N_2}\left(\sum_{N_2 \in \Psi}c_{N_1N_2} - 1\right)} \quad (9)$$

$$d_{N_1N_2} = \begin{cases} \frac{1}{c_{N_1N_2}}, & \text{if } c_{N_1N_2} \neq 0 \\ 1, & \text{if } c_{N_1N_2} = 0 \end{cases} \quad (10)$$

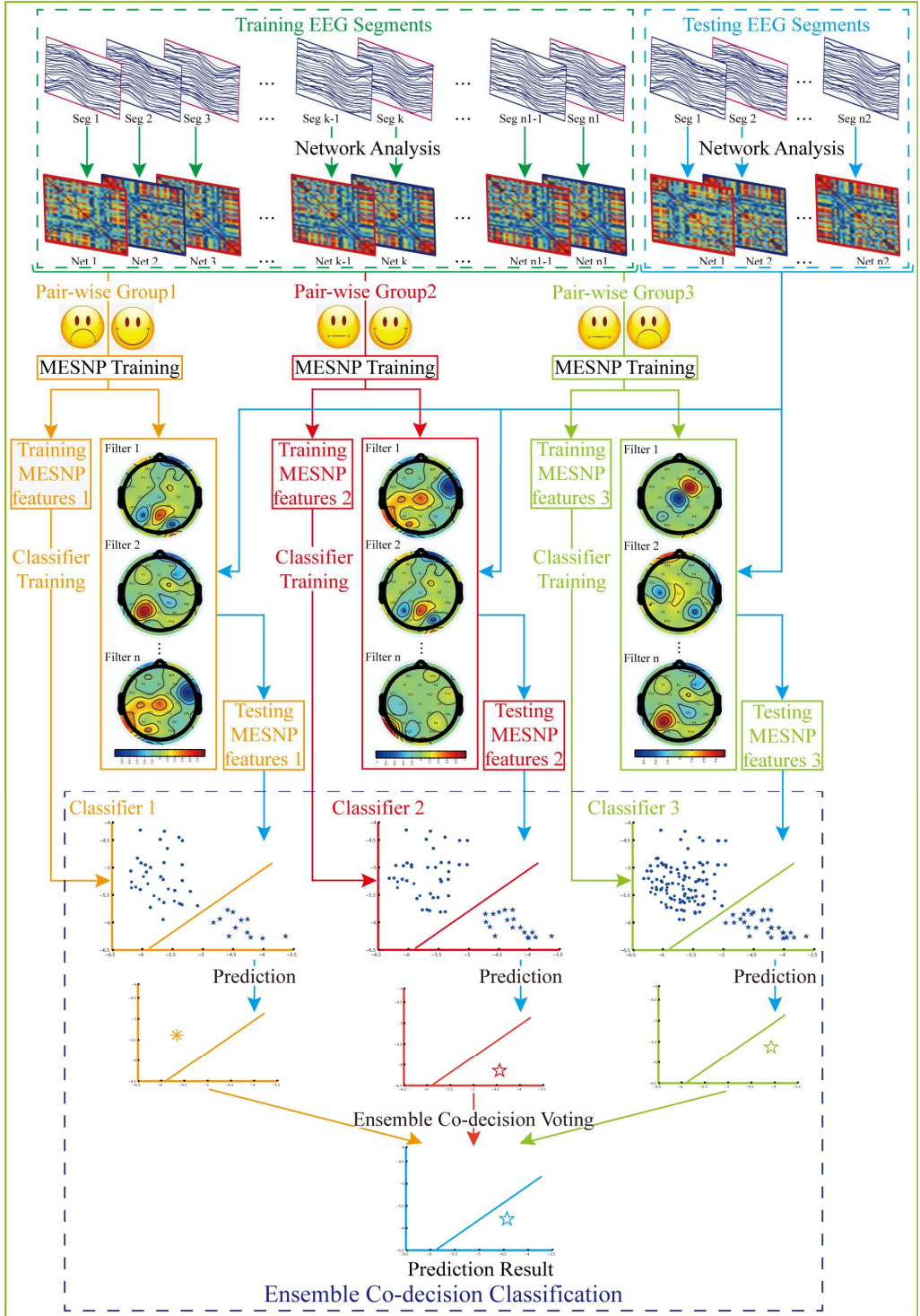

Fig. 1. The overall structure of multiple hierarchical emotional recognition framework.

$$L = \frac{1}{(1/N(N-1))\sum_{N_1 \in \Psi}\sum_{\substack{N_2 \in \Psi \\ N_1 \neq N_2}}(1/d_{N_1N_2})} \quad (11)$$

$$Ge = \frac{1}{N(N-1)}\sum_{N_1 \in \Psi}\sum_{\substack{N_2 \in \Psi \\ N_1 \neq N_2}}(1/d_{N_1N_2}) \quad (12)$$

$$Le = \frac{1}{N}\sum_{N_1 \in \Psi}\frac{\sum_{\substack{N_2,N_3 \in \Psi \\ N_2 \neq N_3}}\left(c_{N_1N_2}c_{N_1N_3}\left[d_{N_2N_3}(\Psi_{N_1})\right]^{-1}\right)^{1/3}}{\sum_{N_2 \in \Psi}c_{N_1N_2}\left(\sum_{N_2 \in \Psi}c_{N_1N_2}-1\right)} \quad (13)$$

In current work, we used these four neural network metrics as features for multiple emotions recognition.

*C. Multiple Emotion-related Spatial Network Topology Patterns for emotion recognition*

Although the network properties can be used to measure the statistical properties of the emotion-related brain neural networks to some degree, the statistical properties may weaken some spatial topology information of the networks and cannot reflect the intact cyberspace information[43]. Considering that the weighted adjacency matrix of the brain neural network contains fundamental spatial information of networks, we fuse supervised learning and ensemble co-decision strategy to extract the discriminative multiple emotion-related spatial network topology patterns (MESNP). The implementation of MESNP consists of two stages, i.e., the training stage and testing stage. During training stage, the parameters for both feature extraction (i.e., spatial filters) and classification are obtained from training data, which will be further utilized for testing data. Fig. 1 illustrates the proposed framework. In essence, the core of MESNP is to learn and identify the discriminative features from the spatial patterns of brain neural networks. In our current work, one-versus-one with max-win strategy is utilized to realize a hierarchal structure for multiple emotions recognition, and the details of the MESNP training and testing implementation are introduced as follows.

*1) Training stage*

Given $\Gamma$ training network matrices $\boldsymbol{G} = \{Plv_1,...,Plv_\Gamma\}$, sized $N \times N$ ($N=32$), which belong to $\kappa$ classes, i.e., $\sum_{i=1}^{\kappa} M_i = \Gamma$, with $i$ representing the $i$-th emotional class and $M_i$ denoting the sample number of the $i$-th emotional class, $Plv$ being the brain neural network sample and $N$ indicating the node number of $Plv$. In essence, the overall implementation of the training processing can be represented as

$$\boxed{Training} \overset{Output}{\Rightarrow} \begin{cases} \boldsymbol{W} = \left\{\underset{\dot{W}_{ij}}{\arg\max} J\left(\dot{W}_{ij}(U_i, U_j)\right)\right\} \\ MESNP_{train} = \left\{\dot{MESNP}_{ij}\right\} \quad \text{with} \begin{cases} i=1,...,(\kappa-1) \\ j=(i+1),...,\kappa \end{cases} \\ Ensem\text{-}Classifier = \{p_{ij}\} \end{cases}$$

(14)

where $\boldsymbol{W}$ represents the pair-coupling spatial filters that trained with supervised method from $\boldsymbol{G} = \{Plv_1,...,Plv_\Gamma\}$, and $MESNP_{train}$ is the corresponding training features. $\boldsymbol{W} = \left\{\underset{\dot{W}_{ij}}{\arg\max} J\left(\dot{W}_{ij}(U_i,U_j)\right)\right\}$ and $MESNP_{train} = \left\{\dot{MESNP}_{ij}\right\}$ are the sub-spatial filters and sub-feature of the training system, with $\dot{W}_{ij}$ indicating the spatial filters for emotion classes $i$ and $j$, $\dot{MESNP}_{ij}$ being the training features for emotion classes $i$ and $j$. *Ensem-Classifier* indicates the ensemble system and $p_{ij}$ denotes the model of pair coupling of binary classification, which is trained based on $\dot{MESNP}_{ij}$ feature in the ensemble system for emotion classes $i$ and $j$. In the following, the details of training sub-spatial filters $\dot{W}_{ij}$, i.e., extracting sub-feature $\dot{MESNP}_{ij}$ and training the sub-classification model $p_{ij}$ of the ensemble system will be introduced.

TABLE I
NOMENCLATURE

| Symbol | Means |
|---|---|
| $X$ | EEG signals |
| $T$ | length of EEG series |
| $F$ | Hamming window function |
| $W_{in}$ | regularization coefficients |
| $f(x)$ | the probability density function of EEG |
| $h$ | differential entropy function |
| $H$ | analysis signal function |
| $A$ | instantaneous amplitude |
| $\varphi$ | instantaneous phase |
| $X_h$ | Hilbert transform |
| $Plv$ | brain neural network |
| $N$ | EEG channel number or network nodes number and $N$=32 |
| $c$ | edge linkage strength |
| $d$ | shortest weighted path length |
| $\Psi$ | set of nodes |
| $\boldsymbol{G}$ | training network set |
| $\Gamma$ | number of training network |
| $\kappa$ | number of classes |
| $M$ | network number of one class |
| $\boldsymbol{W}$ | spatial filters set |
| $W$ | spatial filters |
| $MESNP_{train}$ | training MESNP features |
| *Ensem-Classifier* | ensemble system |
| $p$ | sub-classifier |
| $\boldsymbol{U}$ | paired training network data |
| $U$ | network matrices dataset of one emotional states |
| $b$ | network sample order in one emotional states |
| $\varphi$ | average covariance matrix |
| $J$ | objective function |
| $w$ | eigenvector |
| $\lambda$ | eigenvalue |
| $Plv_{test}$ | testing sample brain neural network |
| $MESNP_{test}$ | testing MESNP features |
| $y$ | sub-classifier prediction |
| $V$ | ensemble voting |
| $k$ | emotional state |
| $Y$ | ensemble system output |

Given a sub-classification task, the spatial filters and binary classifier $p_{ij}$ are trained from two different emotional brain neural networks in the training dataset. Let $\boldsymbol{U}$ represent the paired training EEG set consisting of $i$-th and $j$-th emotional states.

$$\boldsymbol{U} = \{(U_i, U_j)\}, with \begin{cases} i=1,...,(\kappa-1) \\ j=(i+1),...,\kappa \end{cases} \quad (15)$$

where $U_i$ and $U_j$ indicate two different network matrices datasets of the $i$-th and the $j$-th emotional states from training dataset $\boldsymbol{G} = \{Plv_1,...,Plv_\Gamma\}$, which can be described as

$$U_i = \{Plv^{(b_i)}\}, with \begin{cases} b_i = 1,...,M_i \\ \sum_{i=1}^{\kappa} M_i = \Gamma \end{cases} \quad (16)$$

where $i$ represents the $i$-th emotional states, $M_i$ indicates the sample number for $i$-th emotional states and $b_i$ denotes the $b$-th sample in the $i$-th emotional states. Then the average covariance matrix $\phi_i$ and $\phi_j$ of $U_i$ and $U_j$ are calculated as

$$\phi_i = \frac{1}{M_i}\left(\sum_{b_i \in (1,2,3,...,M_i)} \left(Plv_i^{(b_i)}\right)\left(Plv_i^{(b_i)}\right)^T\right) \quad (17)$$

$$\phi_j = \frac{1}{M_j}\left(\sum_{b_j \in (1,2,3,...,M_j)} \left(Plv_j^{(b_j)}\right)\left(Plv_j^{(b_j)}\right)^T\right) \quad (18)$$

In essence, the goal of MESNP is to find the optimal projection to maximize the difference between two different emotional states by maximizing the variance of brain neural networks from one emotional states while minimizing the variance of another [44]. The theoretical implementation is mainly accomplished by computing the diagonalization of the covariance matrix [45]. Actually, the solution of optimal projection problem can be equalized to maximize the following function:

$$\arg_w \max J(w) = \frac{w^T \phi_i w}{w^T \phi_j w} \quad (19)$$

Due to the fact that the scaling of the projection $w$ has no effect on the object value, the optimal projection problem of (19) can be converted to the constrained optimization problem as follows:

$$\begin{cases} \arg_w \max w^T \phi_i w \\ subject\ to\ w^T \phi_j w = 1 \end{cases} \quad (20)$$

Furthermore, introducing the Lagrange multiplier, the constrained optimization problem of (20) can be rewritten as:

$$L(w, \lambda) = w^T \phi_i w - \lambda(w^T \phi_j w - 1) \quad (21)$$

By taking the derivative of (21), under the condition of $\frac{\partial L}{\partial w} = 0$, the projection of $w$, can be estimated with the generalized eigenvalue equation, as follow:

$$\phi_i w = \lambda \phi_j w \quad (22)$$

where $\lambda$ denotes the eigenvalue of generalized eigenvalue equation, and $w$ is the corresponding eigenvector [46]. For multiple spatial filters, equation (22) can be solved as:

$$(\phi_j)^{-1} \phi_i \dot{W} = \sum \dot{W} \quad (23)$$

where $\dot{W}$ consists of the eigenvectors of $(\phi_j)^{-1}\phi_i$, and $\sum = diag(\lambda_1, \lambda_2, ..., \lambda_N)$ is a diagonal matrix with $\lambda$ being the corresponding singular values. In fact, the diagonal values in $\sum$ represent the differential capabilities of spatial filters, and the first and last spatial filters correspond to the largest and smallest eigenvalues which consist of the most discriminative spatial filter pair. In this work, three pairs of spatial filters were adopted to extract MESNP feature for each label pair-wise group. Thus, the trained spatial filters for the $i$-th and $j$-th emotional states can be concatenated as:

$$\dot{W}_{ij} = [Filter_1, Filter_2, Filter_3, Filter_4, Filter_5, Filter_6] \quad (24)$$

where $Filter_k (1 \leq k \leq 6)$ is the eigenvectors in (23). With these filters, the training MESNP features corresponding to the combination of $i$-th and $j$-th emotional states can be extracted as:

$$\dot{MESNP}_{ij}^{(b)} = \log\left(var\left(\dot{W}_{ij}^T Plv_{ij}^{(b)}\right)\right), Plv_{ij}^{(b)} \in \{U_i \cup U_j\} \quad (25)$$

where $Plv_{ij}^{(b)}$ represents the $b$-th brain neural network in the pair-coupling training datasets, with $b=1,2,...,(M_i + M_j)$, and $var(\cdot)$ represents the variance of each spatial filtered data, resulting in a 6-length vector when 3 pairs of filters are used. Obviously, the spatial filters $\dot{W}_{ij}$ is trained with the supervised learning strategy, which results in the label guided spatial features. Thus, the training MESNP feature set for the combination of $(U_i, U_j)$ can be defined as

$$\dot{MESNP}_{ij} = \left\{\dot{MESNP}_{ij}^{(1)}, \dot{MESNP}_{ij}^{(2)}, ..., \dot{MESNP}_{ij}^{(M_i+M_j)}\right\} \quad (26)$$

Where each element in (26) denotes the MESNP feature of one sample in set $\{U_i \cup U_j\}$. Based on the extracted spatial features $\dot{MESNP}_{ij}$, the sub-classification model $p_{ij}$ of the ensemble system corresponding to set $\{U_i \cup U_j\}$ can be trained. In this work, the SVM with 5-fold cross validation based parameter optimization[47] is utilized as the sub-classifier.

Due to the fact that the result of pair-coupling for training dataset includes $\kappa(\kappa-1)/2$ groups, the training results should contains $\kappa(\kappa-1)/2$ sets of spatial filters and MESNP features, respectively. Finally, the trained pair-coupling spatial filters and their corresponding features can be integrated as

$$W = \{\dot{W}_{ij}\}, with \begin{cases} i = 1,...,(\kappa-1); \\ j = (i+1),...,\kappa \end{cases} \quad (27)$$

$$MESNP_{train} = \{\dot{MESNP}_{ij}\}, with \begin{cases} i = 1,...,(\kappa-1); \\ j = (i+1),...,\kappa \end{cases} \quad (28)$$

Obviously, corresponding to the training groups, totally $\kappa(\kappa-1)/2$ sub-classifiers need to be further trained for the ensemble system so as that the system can be used for multi-class identification tasks. The ensemble system is defined as

$$Ensem\text{-}Classifier = \{P_{ij}\}, with \begin{cases} i = 1,...,(\kappa-1); \\ j = (i+1),...,\kappa \end{cases} \quad (29)$$

At this point, the training procedure already has been accomplished to get the spatial filter sets $W = \{\dot{W}_{ij}\}$, training MESNP features $MESNP_{train}$ and the ensemble system $Ensem\text{-}Classifier$. The training procedure can be summarized in Algorithm 2.

**Algorithm 2** Training process

1: **Input**: Training dataset $G = \{Plv_1,...,Plv_\Gamma\}$
2: $\Gamma \rightarrow$ number of training network matrices
3: $Plv \rightarrow$ the brain neural network sample, sized $32 \times 32$
4: $\kappa \rightarrow$ number of training emotional label categories
5: $M_i \rightarrow$ number of $i$-th emotional state brain neural networks and
$\sum_{i=1}^{\kappa} M_i = \Gamma$
6: **Output**: $W \rightarrow$ spatial filter sets
  $MESNP_{train} \rightarrow$ training MESNP feature
  $Ensem\text{-}Classifier \rightarrow$ ensemble system
7: **Initialization**:
8: for $i \rightarrow 1,...,(\kappa-1)$
9:   for $j \rightarrow (i+1),...,\kappa$
10:     $U_i = \{Plv^{(b_i)}\}, b_i = 1,...,M_i$ and
$U_j = \{Plv^{(b_j)}\}, b_j = 1,...,M_j$
11:     training spatial filter $\dot{W}_{ij}$ and extracting training MESNP feature
$\dot{MESNP}_{ij}$ with $(U_i, U_j)$
12:     training sub-model $P_{ij}$ of ensemble system with $\dot{MESNP}_{ij}$
13:     $W = \{\dot{W}_{ij}\} \xleftarrow{update} \dot{W}_{ij}$
14:     $MESNP_{train} = \{\dot{MESNP}_{ij}\} \xleftarrow{update} \dot{MESNP}_{ij}$
15:     $Ensem\text{-}Classifier = \{P_{ij}\} \xleftarrow{update} P_{ij}$
16:   end
17: end

*2) Testing stage*

During testing stage, for a given testing brain neural network $Plv_{test}$, its MESNP features can be extracted with the trained spatial filters set $W$, which shown in (27), and the testing MESNP features can be computed as

$$MESNP_{test} = \log\left(\text{var}\left(W^T Plv_{test}\right)\right) \quad (30)$$

As shown in (27), due to the fact that there are $\kappa(\kappa-1)/2$ pair-coupling for training dataset and $W$ includes $\kappa(\kappa-1)/2$ sets of spatial filters, the testing MESNP features should contain $\kappa(\kappa-1)/2$ MESNP feature groups. The testing MESNP feature group can be estimated as

$$\dot{MESNP}_{ij}^{(test)} = \log\left(\text{var}\left(\dot{W}_{ij}^T Plv_{test}\right)\right), with \begin{cases} i = 1,...,(\kappa-1); \\ j = (i+1),...,\kappa \end{cases} \quad (31)$$

where $W_{ij}$ corresponds to the pair-coupling spatial filters, which trained from the training sub-classification task with the pair-coupling of $i$-th and $j$-th emotional states, and the testing MESNP feature $\dot{MESNP}_{ij}^{(test)}$ will be predicted with the sub-classification model $p_{ij}$ in the prediction process.

Obviously, for each testing network sample, there should be $\kappa(\kappa-1)/2$ MESNP feature groups. Consequently, the final MESNP features for each testing sample $Plv_{test}$ is as

$$MESNP_{test} = \{\dot{MESNP}_{ij}^{(test)}\}, with \begin{cases} i = 1,...,(\kappa-1); \\ j = (i+1),...,\kappa \end{cases} \quad (32)$$

The extracted MESNP features for testing sample in (32) will be sequentially inputted into the corresponding trained sub-classifiers $P_{ij}$, resulting in $\kappa(\kappa-1)/2$ label predicting outcomes, which denoted as $y_{ij}$, from *Ensem-Classifier*. Thus, the predicting of the sub-classifiers $P_{ij}$ with testing MESNP features $\dot{MESNP}_{ij}^{(test)}$ can be estimated as

$$P_{ij}\left(\dot{MESNP}_{ij}^{(test)}\right) \xRightarrow{output} y_{ij} = \begin{cases} i, \text{ if } y_{ij} \in Emotion_i \\ j, \text{ if } y_{ij} \in Emotion_j \end{cases} \quad (33)$$

As (33) shows, the $MESNP_{test}$ will be put into the ensemble system $EnsemClassifier = \{P_{ij}\}$, and each testing MESNP feature will be predicted by the corresponding sub-classification SVM model already trained with features extracted from the same spatial filters. After predicted from each sub-classification SVM model of the ensemble system, the final predicting emotion label will be predicted by ensemble co-decision voting as

$$\begin{cases} V = \{V_k\} \\ V_k = \sum_{i=1}^{(\kappa-1)} \sum_{j=i+1}^{\kappa} 1 | y_{ij} == k \end{cases}, with\ k = 1,2,...,\kappa \quad (34)$$

where $V_k$ indicates the frequency of the $k$-th class identified by the ensemble system, i.e. the vote of the ensemble co-decision voting, and $V$ is the voting results set of all

**Algorithm 3** Testing process

1: **Input**: Testing brain neural network $Plv_{test}$
2: **Output**: Predicting result
3: for $i \rightarrow 1,...,(\kappa-1)$
4:   for $j \rightarrow i+1,...,\kappa$
5:     $\dot{MESNP}_{ij}^{(test)} = \log\left(\text{var}\left(\dot{W}_{ij}^T Plv_{test}\right)\right)$
6:     $MESNP_{test} = \{\dot{MESNP}_{ij}^{(test)}\} \xleftarrow{update} \dot{MESNP}_{ij}^{(test)}$
7:     $P_{ij}\left(\dot{MESNP}_{ij}^{(test)}\right) \xRightarrow{output} y_{ij}$
8:   end
9: end
10: for $k \rightarrow 1,2,...,\kappa$
11:   $V_k = \sum_{i=1}^{(\kappa-1)} \sum_{j=i+1}^{\kappa} 1 | y_{ij} == k$
12:   $V = \{V_k\} \xleftarrow{update} V_k$
13: end
14: $Y_{test} = \{k | \max_k [V_1, V_2,...,V_k,...,V_\kappa]\} \Rightarrow (Y_{test} = k)$

emotional states for current testing sample. Ideally, the final output of testing emotional label for the *k*-th class can be expressed as

$$Y_{test} = \{k \mid \max_k [V_1, V_2, ..., V_k, ..., V_K]\} \Rightarrow (Y_{test} = k) \quad (35)$$

As shown in (35), the final label prediction result for the testing brain neural network $Plv_{test}$ is the class having the highest voting value in the ensemble co-decision system. Fig.1 demonstrates the implementation procedure of MESNP for MAHNOB-HCI dataset, which includes three emotional states, i.e. Negative, Positive and Neutral. Similar to MAHNOB-HCI, the MESNP of DEAP includes six pair-wise groups and the ensemble system includes six sub-models, and the final prediction for testing sample will be decided from the voting of six sub-models prediction. The testing procedure can be summarized in Algorithm 3.

## IV. MATERIALS

In this work, two public emotional datasets are utilized, i.e., MAHNOB-HCI and DEAP. Videos were used as experimental stimulus for these two datasets. When participants watching these stimuli, the EEG-acquisition equipment will record the emotion-related EEG signals. MAHNOB-HCI dataset consists of three emotion categories, while DEAP dataset has 4 emotion states evoked.

### A. MAHNOB-HCI Dataset

The MAHNOB-HCI dataset contains EEG, peripheral physiological signals, functional near-infrared spectroscopy (fNIRS) and facial videos of 27 participants (11 male and 16 female, aged between 19 and 40). During experiments, 32-channel electrodes were assigned on the participants' scalp in accordance with the international standard 10-20 system to collect EEG data and the sampling frequency is 256 Hz. The experimental stimuli videos were selected by participants via an on-line self-assessment manikins (SAM) system from 155 video clips. Through statistical assessment of participants' self-emotion, 20 video segments from 155 video segments were selected as the stimuli of the experiment [33]. In the experiment, 20 video segments were played randomly and the participants had to use some emotional labels to express their emotions when they watched the video, the labels include neutral, anxiety, amusement, sadness, joy, disgust, anger, surprise, fear, happiness and so on. The labels were divided into three categories from the dimensions of Arousal and Valence, including positive, neutral and negative emotions, and the specific division is shown in Table II. The detail of MAHNOB-HCI dataset could refer to [33].

TABLE II
THE SPECIFIC DIVISION OF EMOTION LABELS FOR MAHNOB-HCI DATASET

| Emotion | Label |
|---|---|
| Positive | Joy, happiness, amusement |
| Neutral | Surprise, neutral |
| Negative | Fear, anger, disgust, sadness, anxiety |

### B. DEAP Dataset

The DEAP database contains EEG and peripheral physiological signals of 32 healthy participants (16 male and 16 females, aged between 19 and 37). The experiment used 48-channel electrodes (32 EEG channels, 12 peripheral channels, 3 unused channels, and 1 status channel) to collect data, and the sampling rate is 512HZ. The EEG channels were placed according to standard 10-20 system. This experiment first selected 120 initial stimuli, half of which were chosen semi-automatically and the rest manually, and then used a web-based subjective emotion assessment interface to choose 40 test video clips as the stimuli. Stimuli induce emotions in the four quadrants of the valence-arousal (VA) space (LALV, HALV, LAHV, and HAHV) as shown in Fig.2. The EEG signals of 32 participants were recorded as each watched 40 one-minute long excerpts of music videos. At the end of each trial, participants performed a self-assessment of their levels of arousal, valence, liking, and dominance [34].

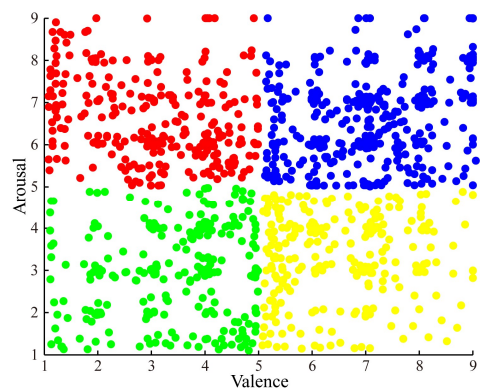
Fig. 2. The valence-arousal (VA) space model

### C. Data processing

To reduce the noise influence and enhance the stability of EEG data, we performed following four steps for EEG preprocessing: 1) segmenting EEG dataset with 10-s moving window; 2) converting segmented EEG data to the average reference; 3) adopting baseline correction to reduce the impact of baseline drift by using the first 1 second signals in each segment; 4) filtering each EEG segment into in four frequency bands with band-pass filters, i.e., 4-8 HZ (theta band), 8-12 HZ (alpha band), 12-30 HZ (beta band), and 30-48 HZ (gamma band).

In the feature extraction stage, we firstly divided EEG segments into training and testing datasets and then constructed brain neural networks with PLV from segmented EEG data. Based on training datasets, we trained MESNP filters and the corresponding ensemble classifiers. In order to verify the effectiveness and feasibility of our method, we also used the widely adopted features, like power spectral density (PSD), differential entropy (DE) and the network properties features as benchmarks for comparison.

### D. DEAP Dataset

In this experiment, the brain neural networks are constructed for all segments. To evaluate the efficiency of our proposed emotional recognition strategy, we designed two classification tasks, i.e., the off-line classification task and the simulated on-line classification task. To simulate the condition of on-line data collection and processing, the preprocessed EEG data were divided into two parts, where the first 50% of the time series was used for off-line analysis to train the model and the last 50% of the time series was used as on-line real-time data acquisition simulation. Specifically, the MESNP filters and the ensemble co-decision classification system are trained in the off-line analysis. Subsequently, the on-line MESNP features are extracted from brain neural networks with the trained MESNP spatial filters, and the final on-line emotional states will be predicted from the ensemble co-decision classification system.

In fact, the main difference between off-line and on-line classifications rises from the fact that whether the temporal sequence information is considered or not. For off-line classification task, the order of time sequence in EEG data is not considered and the 10-fold cross-validation scheme is used to randomly divide the EEG segments into training and testing sets. To get the robust and convinced result, the 10-fold cross-validation is repeated for 10 times for each classification approach, and the mean accuracy across the 10 times is reported. For on-line classification task, the sequential order of EEG data must be considered to simulate the situation of on-line real-time data acquisition (i.e., those segments recorded in the early stage will be used as training samples, while other segments recorded in the relatively later stage will be served as the testing samples).

## V. EXPERIMENT RESULTS

### A. Experiment results on MAHNOB-HCI Dataset

For the MAHNOB-HCI Dataset, three classes of emotions (positive, neutral and negative) were labeled in the whole experiment. Based on the four frequency bands (i.e., theta, alpha, beta and gamma), respectively, we utilized Network properties, PSD, DE and MESNP features to perform the prediction of the three emotions. The corresponding classification accuracies are shown in Table III and Fig.3, respectively.

TABLE III
THE EMOTION CLASSIFICATION ACCURACIES (%) ON MAHNOB-HCI DATASET

| Band / Features | theta | alpha | beta | gamma |
|---|---|---|---|---|
| Network Properties | 51.79±9.35 | 53.45±8.70 | 50.88±10.24 | **55.21±11.41** |
| PSD | 52.14±7.29 | 52.67±7.68 | **60.62±7.76** | 58.61±9.79 |
| DE | 52.47±7.37 | 53.80±7.34 | 64.56±7.38 | **71.25±7.28** |
| MESNP | **99.93±0.13** | 99.85±0.42 | 98.41±1.84 | 96.20±2.60 |

As shown in Table III, we can find that the optimal recognition rates are 55.21%(gamma), 60.62% (beta) and 71.25% (gamma) for the Network Properties features, PSD and DE features in all the four frequency bands, respectively. For MESNP features, however, the prediction accuracy has achieved 99.93%±0.13, 99.85%±0.42, 98.41%±1.84 and 96.20%±2.60 in theta, alpha, beta and gamma bands, respectively. In addition, the results in lower frequency bands (i.e., theta and alpha) are better than that in the higher frequency bands (i.e., beta and gamma).

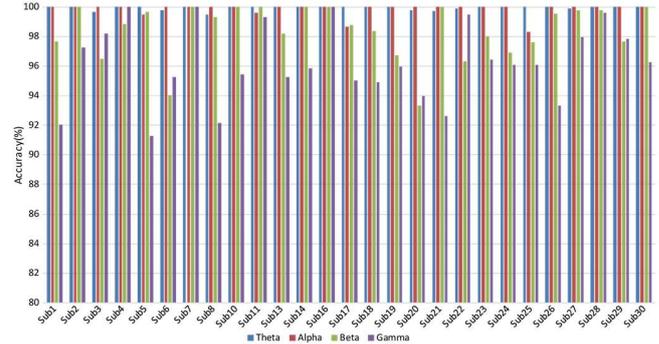

Fig. 3. The classification accuracies with MESNP features for each participant on MAHNOB-HCI dataset

Table III clearly shows that the recognition based on MESNP features are better than other features widely used for emotion classification in previous studies. Based on MESNP features, we further explored the prediction performances across all participants in the four frequency bands (Fig.3). Fig.3 presents the variability across participants (i.e., 99.47%-100% for the theta band; 98.30%-100% for the alpha band; 93.32%-100% for the beta band; 91.26%-100% for the gamma band). Importantly, all the participants had the prediction accuracies over 90%, and the best prediction performance was also exhibited in the low frequency bands (theta and alpha), which was consistent with the result in Table III.

We further compared the recognition accuracy of various systems using MAHNOB-HCI dataset and presented in Table IV.

TABLE IV
COMPARISON OF VARIOUS STUDIES ON MAHNOB-HCI DATASET

| Study | Results |
|---|---|
| Koelstra et al.[2] | Average classification rates of 80.00%, 80.00% and 86.70% for arousal, valence, and control ratings (2 classes) with 24 participants. |
| Huang et al.[4] | The best classification rate (66.28% for valence and 63.22% for arousal) (2 classes) is achieved. |
| Wang et al. [7] | For valence and arousal (2 classes), the classification rates are 61.35% (arousal) and 60.22% (valence). |
| Our method | Average classification rate of 99.93% for 3 classes (positive, neutral and negative) with all 27 participants. |

Koelstra et al. [2] presented a multi-modal approach that analyzed both facial expressions and electroencephalography (EEG) signals for the generation of affective tags. They performed binary classification on the arousal, valence and control ratings, which are threshold into high (rating 6-9) and low (rating 1-5) classes. For arousal, valence, and control, video tag classification rates of 80.00%, 80.00%, and 86.70% are obtained respectively when aggregating across all participants. Huang et al. fused the facial expression features and EEG features for emotion recognition and the best classification rate for valence and arousal are 66.28% and 63.22% for 2 classes. Wang et al. [7]proposed a novel emotion recognition approach with privileged information by exploiting relations between EEG signals and stimulus videos. The best accuracies for arousal and valence (2 classes) are 61.35% and 61.22%. Most of those reported studies focused on the 2-classes emotion recognition. Our method for classification of 3 classes (positive, neutral and negative) has achieved the average accuracy of 99.93% on the same dataset.

The average on-line simulation experimental recognition result of all participants for MAHNOB-HCI dataset is shown in Fig.4. We can find that the experimental results in theta and alpha bands have achieved 100% for every participant. Though the results in beta and gamma are less stable than that in theta and alpha bands, the average classification results are still above 99% in both bands. Combined with the off-line experimental results, we could discover that the results of two experiments are consistent and the better recognition results are both in theta and alpha bands for this dataset.

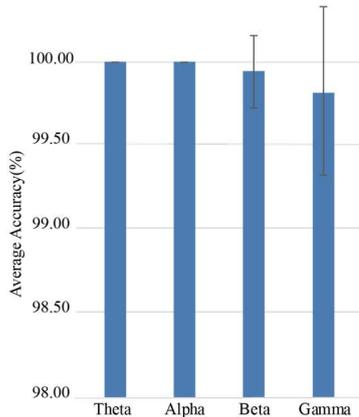

Fig. 4. The on-line simulation classification accuracy for MAHNOB-HCI dataset

### B. Experiment results on DEAP Dataset

For DEAP Dataset, we divided the valence-arousal (VA) space into four quadrants (rating 1-5 is defined as low, rating 5-9 is defined as high, resulting in LALV, LAHV, HALV, HAHV 4 classes, shown in Fig. 3). The experimental results for different features of all 32 participants are shown in Table V and Fig.5.

TABLE V
THE EMOTION CLASSIFICATION ACCURACIES (%) ON DEAP DATASET

| Band<br>Features | theta | alpha | beta | gamma |
|---|---|---|---|---|
| Network Properties | 64.28±5.22 | **73.33±4.47** | 41.31±7.74 | 42.84±6.48 |
| PSD | 47.69±6.70 | 50.57±7.05 | **57.57±6.59** | 47.21±6.57 |
| DE | 46.92±6.76 | 50.04±6.39 | **55.56±6.47** | 41.83±7.27 |
| MESNP | 81.81±3.17 | **83.66±2.43** | 36.64±5.94 | 42.75±6.02 |

As shown in Table V, the best recognition rates are 73.33% ±4.47 (alpha), 57.57% ±6.59 (beta) and 55.56%±6.47 (beta) for the Network Properties features, PSD and DE features in all four frequency bands, respectively. As for MESNP features, the best accuracy rate was 83.66% ± 2.43 (alpha), which is significantly higher than other features (Network Properties features, PSD and DE).

The recognition accuracy for every participant is shown in Fig.5. The accuracies of low bands (theta and alpha) apparently are better than that of high bands (beta and gamma) for MESNP based approach. In Table V, the average accuracy also shows the obvious disparities among different bands, where the average accuracies reach up to 81.81% and 83.66% in theta and alpha bands, however, 36.64% and 42.78% in beta and gamma bands. Comparing Table III with Table V, the classification results of DEAP are highly consistent with that of MAHNOB-HCI, which could verify our proposed method to be efficient and stable for emotion recognition. We compared our work with previously reported studies for DEAP dataset as shown in Table VI.

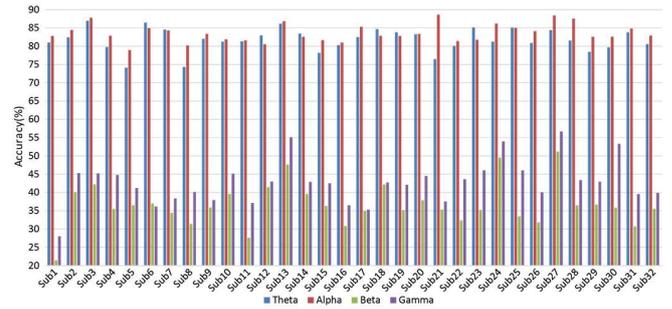

Fig. 5. The classification accuracies with MESNP features for each participant on DEAP dataset

TABLE VI
COMPARISON OF VARIOUS STUDIES ON DEAP DATASET

| Study | Results |
|---|---|
| Yoon et al. [1] | The classification rates of 70.90%, 70.10% for valence and arousal (2 classes), 55.40%, 55.20% for valence and arousal (3 classes) with all 32 participants. |
| Liu et al. [3] | Selected 10 participants and achieved 63.04% for arousal-dominance recognition (4 classes). |
| Zhang et al. [6] | Chose 8 participants and achieved 75.19 % and 81.74 % on valence and arousal (2 classes). |
| Zheng et al.[9] | 69.67% for quadrants of VA space (4 classes) with all 32 participants. |
| Arnau-González et al.[11] | 67.70% and 69.90% on arousal and valence (2 classes) with all 32 participants. |
| Our method | Average classification rates of 83.66% on valence-arousal (VA) space (4 classes) with all 32 participants. |

Yoon et al. [1] defined a probabilistic classifier based on Bayes' theorem for valence and arousal classification with DEAP dataset. The best classification rates for their experiment are 70.90% (2 classes) and 55.40% (3 classes). Liu et al. [3] proposed a real-time fractal dimension (FD) based valence level recognition algorithm for EEG signals. In this study, they selected 10 participants from DEAP dataset as the material for their experiments and reached the best mean accuracy of 63.04% for arousal-dominance recognition (4 classes). Zhang et al. [6] designed an ontological model to represent and integrate EEG data, which achieved an average recognition rate of 75.19% on valence and 81.74% on arousal (2 classes). Zheng et al. [9] used a newly developed pattern classifier named discriminative Graph regularized Extreme Learning Machine (GELM) and extracted the differential entropy (DE) as their training feature. Their method achieved an average accuracy of 69.67% on the DEAP dataset for quadrants of VA space (4 classes) in theta frequency band for all participants. Arnau-González et al. [11] combined both connectivity-based and channel-based features with a selection method and achieved 67.70% (arousal) and 69.90% (valence) on arousal and valence (2 classes). Compared with their work, we have the largest number of categories (4 classes) and the highest recognition accuracy (83.66%), which highlights the superiority and stability of our method.

Fig.6 shows the on-line simulation experimental recognition of all participants on DEAP dataset. The similar results to MAHNOB-HCI dataset can be found in DEAP dataset, where the results in theta and alpha bands are more stable than those in beta and gamma bands. The best on-line experimental classification result on DEAP dataset has achieved 99.22% in the alpha band, which is consistent with the off-line classification result.

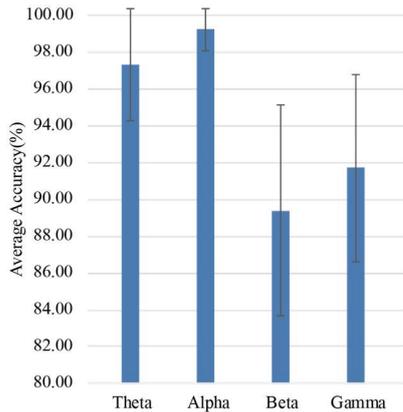

Fig. 6. The on-line simulation classification accuracy for DEAP dataset

Moreover, in order to investigate how MESNP can effectively extract the discriminative spatial network patterns, we performed the analysis of variance (ANOVA) to explore the possible relationship between the learned spatial filters and network patterns of emotion states. The statistically significant difference edges ($p<0.01$) between different emotion brain neural networks and the learned two most discriminative spatial filters in alpha band for one subject in MAHNOB-HCI dataset are shown in Fig.7, where (a), (b) and (c) correspond to positive vs negative, positive vs neutral and negative vs neutral, respectively. Specifically investigating Fig.7, we could find that the different emotion pairs exhibited the different distinct network patterns. The proposed MESNP can adaptively learn the specific discriminative spatial network patterns, where the vital network nodes of spatial filters are emphasized with large values (i.e., those marked with either the red or deep blue colors), while other less important nodes are compressed by giving the small weight values.

## VI. DISCUSSION

The challenge for the reliable classification of different emotions is mainly due to the very limited knowledge regarding the underlying neural mechanism of emotions. In this work, we mainly investigate the feature extraction and classification based on emotion-related EEG brain neural networks that can measure the information propragration and exchange among different brain areas. Considering that different emotional processing involves different brain network patterns, this work established an approach termed as MESNP to extract EEG-related brain neural networks topology features of different emotional states. We specifically constructed the emotion-related brain neural networks based on EEG data, and then combined the supervised learning and the ensemble co-decision together to train the spatial filters to extract spatial topology difference features from brain neural networks, and also train the co-decision classification system.

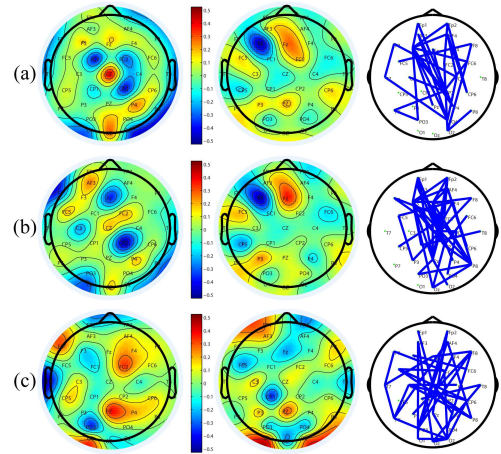

Fig. 7. The scalp topologies for the two most discriminative MESNP filters (Filters 1 and 6) and the topology difference in the brain neural networks between different emotional states in the alpha band for one subject in MAHNOB-HCI dataset.

In this work, based on the public emotional MAHNOB-HCI and DEAP datasets, we performed the off-line and on-line simulation experiments to investigate the feasibility of proposed method. From the experiment results of Table III and Table V, we can find that MESNP can achieve the higher classification accuracy than the other conventional approaches. Moreover, MESNP consistently reveals the relatively better performances in the lower frequency bands (theta and alpha) compared to other two higher frequency bands (beta and gamma) for both two emotion datasets. The performance improvement of MESNP infers that the MESNP features extracted from brain neural networks may contain more discriminative information to differentiate the human emotions. Combining the network differences between two paired emotional states and the corresponding spatial MNESP filters in Fig.7, the working mechanism of MNESP could be revealed.

Fig.7 shows that there exists network topology difference between brain neural networks (positive, neutral and negative) for different emotions, which means that the spatial discriminative information exists in the network topologies. Comparing the differential network topological patterns with the spatial filters, we can discover that the nodes exhibiting large topology differences are imposed with the larger filter weights, while other less important nodes are given with the smaller filter weights. Essentially, when multiplying the filters with the network adjacency matrix, the spatial filters function like a sub-network-seeking filters, automatically selecting modules with significant spatial differences between emotions in the brain neural network space by giving them larger emphasis and compressing other modules by providing smaller coefficients. Therefore, this band-pass-like filter can be used to effectively extract the spatial topological differences between different emotions, so as to obtain more reliable distinctions.

Specifically, investigating Fig.7 we can also find that though the network topology difference mainly exists in the frontal, occipital and parietal regions that are proved to be associated with emotions [48-52], the different emotion pairs actually show the distinct discriminative network patterns. Due to the adoption of the supervised learning strategy, MNESP can adaptively capture these differences, resulting in different spatial MNESP patterns for different emotion pairs. Besides the differences in network topology analysis in alpha band as reported in Fig.7, we also performed the same analysis for other three frequency bands. The comparison among the four frequency bands reveals that there are significant differences between different emotions network in the theta and alpha bands of two datasets while the differences in the beta and gamma bands are relatively weak, which accounts for the highest classification results obtained in the theta and alpha bands. Previous studies also confirmed that the more discriminative information for emotion recognition are revealed in the theta and alpha bands. Lin et al. used SVM to classify emotions and the results of asymmetry feature in alpha and theta bands have achieved the highest accuracy among all four frequency bands [25]. Shahabi et al. found that the frontal-to-parietal connectivity in the alpha band specifically increased when emotions changed from neutral to joy and in almost every case, the meaningful differences were observed in theta band for emotion shift from melancholic to joy [53]. In the theta band, compared to happiness emotion, significant higher coherence is found in healthy participants during sadness, fear, disgust, and anger emotion states, and in the alpha band, sadness and disgust appeared to have higher coherence than happiness and surprise emotions [54]. Moreover, the asymmetries at the F3-F4 pair related to valence emotion are observed in both alpha and theta bands when analyzing the emotional arousal during affective-pictures stimuli [48, 49].

Besides the conventional fold cross-validation based evaluation, this work newly performed a simulated on-line analysis, which is much closer to the practical application. Similar to the off-line analysis, the analysis for the simulated on-line protocol also achieves the relatively high accuracy above 88%, which may provide the promising tool for the realization of the effective on-line affective recognition system. In the filed of BCI, how to improve the emotional interaction between computers and humans is the biggest chalenge for researchers. The simulated on-line experimental results also shows the stability and effectiveness of the framework we proposed, and this framework may provide a possibility for the realization of intelligent affective brain computer interfaces system.

## VII. CONCLUSION

In the current study, based on the public datasets, we have systematically compared the performance differences between conventional features and the discriminative network features proposed in this work for emotional recognition. The results in both off-line and on-line classification tasks showed that the proposed feature extraction method MESNP can robustly and reliably differentiate different emotional states, and the MESNP features can be used to make better categorical predictions of emotions.

ACKNOWLEDGMENT

We would like to thank the databases (MAHNOB-HCI, DEAP) providers here, and thanks all the participants in the emotion experiments of these two datasets.

REFERENCES

[1] H. J. Yoon, and S. Y. Chung, "EEG-based emotion estimation using Bayesian weighted-log-posterior function and perceptron convergence algorithm," *Computers in Biology & Medicine,* vol. 43, no. 12, pp. 2230-2237, 2013.
[2] S. Koelstra, and I. Patras, "Fusion of facial expressions and EEG for implicit affective tagging," *Image and Vision Computing,* vol. 31, no. 2, pp. 11, 2013.
[3] Y. Liu, and O. Sourina, "Real-time fractal-based valence level recognition from EEG," *Transactions on computational science XVIII,* pp. 101-120: Springer, 2013.
[4] X. Huang, J. Kortelainen, G. Zhao, X. Li, A. Moilanen, T. Seppänen, and M. Pietikäinen, "Multi-modal emotion analysis from facial expressions and electroencephalogram," *Computer Vision and Image Understanding,* vol. 147, pp. 114-124, 2016.
[5] S. Spence, "Descartes' Error: Emotion, Reason and the Human Brain," *Bmj Clinical Research,* vol. 310, no. 6988, pp. 1213-1213, 1995.
[6] Zhang, Xiaowei, Hu, Bin, Chen, Jing, Moore, and Philip, "Ontology-based context modeling for emotion recognition in an;intelligent web," *World Wide Web-internet & Web Information Systems,* vol. 16, no. 4, pp. 497-513, 2013.
[7] S. Wang, Y. Zhu, L. Yue, and J. Qiang, "Emotion Recognition with the Help of Privileged Information," *IEEE Transactions on Autonomous Mental Development,* vol. 7, no. 3, pp. 189-200, 2015.
[8] P. Salovey, and J. D. Mayer, "Emotional intelligence," *Imagination, cognition and personality,* vol. 9, no. 3, pp. 185-211, 1990.
[9] W. L. Zheng, J. Y. Zhu, and B. L. Lu, "Identifying Stable Patterns over Time for Emotion Recognition from EEG," *IEEE Transactions on Affective Computing,* vol. PP, no. 99, pp. 1-1, 2017.
[10] W.-L. Zheng, W. Liu, Y. Lu, B.-L. Lu, and A. Cichocki, "Emotionmeter: A multimodal framework for recognizing human emotions," *IEEE transactions on cybernetics,* no. 99, pp. 1-13, 2018.
[11] P. Arnau-González, M. Arevalillo-Herráez, and N. Ramzan, "Fusing highly dimensional energy and connectivity features to identify affective states from EEG signals," *Neurocomputing,* vol. 244, pp. 81-89, 2017.
[12] C. Mühl, B. Allison, A. Nijholt, and G. Chanel, "A survey of affective brain computer interfaces: principles, state-of-the-art, and challenges," *Brain-Computer Interfaces,* vol. 1, no. 2, pp. 66-84, 2014.
[13] A. Al-Nafjan, M. Hosny, Y. Al-Ohali, and A. Al-Wabil, "Review and classification of emotion recognition based on EEG brain-computer interface system research: a systematic review," *Applied Sciences,* vol. 7, no. 12, pp. 1239, 2017.
[14] H. Huang, Q. Xie, J. Pan, Y. He, Z. Wen, R. Yu, and Y. Li, "An EEG-Based Brain Computer Interface for Emotion Recognition and Its Application in Patients with Disorder of Consciousness," *IEEE Transactions on Affective Computing,* 2019.
[15] J. Atkinson, and D. Campos, "Improving BCI-based emotion recognition by combining EEG feature selection and kernel classifiers," *Expert Systems with Applications,* vol. 47, pp. 35-41, 2016.
[16] E. T. Esfahani, and V. Sundararajan, "Using brain-computer interfaces to detect human satisfaction in human-robot interaction," *International Journal of Humanoid Robotics,* vol. 8, no. 01, pp. 87-101, 2011.
[17] R. W. Levenson, "Human emotion: A functional view.," *The nature of emotion: Fundamental questions,* vol. 1, pp. 123-126, 1994.
[18] M. B. Arnold, and J. Gasson, "Feelings and emotions as dynamic factors in personality integration," *The human person,* pp. 294-313, 1954.

[19] X.-W. Wang, D. Nie, and B.-L. Lu, "Emotional state classification from EEG data using machine learning approach," *Neurocomputing,* vol. 129, pp. 94-106, 2014.

[20] R. Jenke, A. Peer, and M. Buss, "Feature Extraction and Selection for Emotion Recognition from EEG," *IEEE Transactions on Affective Computing,* vol. 5, no. 3, pp. 327-339, 2017.

[21] K. Jonghwa, and A. Elisabeth, "Emotion recognition based on physiological changes in music listening," *IEEE Transactions on Pattern Analysis & Machine Intelligence,* vol. 30, no. 12, pp. 2067-2083, 2008.

[22] Z. Zhihong, P. Maja, R. Glenn I, and H. Thomas S, "A survey of affect recognition methods: audio, visual, and spontaneous expressions," *IEEE Transactions on Pattern Analysis and Machine Intelligence,* vol. 31, no. 1, pp. 39-58, 2009.

[23] R. Munoz, R. Olivares, C. Taramasco, R. Villarroel, R. Soto, T. S. Barcelos, E. Merino, and M. F. Alonso-Sánchez, "Using Black Hole Algorithm to Improve EEG-Based Emotion Recognition," *Computational Intelligence and Neuroscience*, 2018.

[24] Y. Dasdemir, E. Yildirim, and S. Yildirim, "Analysis of functional brain connections for positive–negative emotions using phase locking value," *Cognitive neurodynamics,* vol. 11, no. 6, pp. 487-500, 2017.

[25] L. Yuan-Pin, W. Chi-Hong, J. Tzyy-Ping, W. Tien-Lin, J. Shyh-Kang, D. Jeng-Ren, and C. Jyh-Horng, "EEG-based emotion recognition in music listening," *IEEE Transactions on Biomedical Engineering,* vol. 57, no. 7, pp. 1798-1806, 2010.

[26] G. Zhao, G. Yan, B. Shen, X. Wei, and W. Hao, "Emotion Analysis for Personality Inference from EEG Signals," *IEEE Transactions on Affective Computing,* vol. 9, no. 3, pp. 362-371, 2017.

[27] S. L. Bressler, and V. Menon, "Large-scale brain networks in cognition: emerging methods and principles," *Trends in cognitive sciences,* vol. 14, no. 6, pp. 277-290, 2010.

[28] L. Liu, L.-L. Zeng, Y. Li, Q. Ma, L. Li, H. Shen, and D. Hu, "Altered cerebellar functional connectivity with intrinsic connectivity networks in adults with major depressive disorder," *PloS one,* vol. 7, no. 6, pp. e39516, 2012.

[29] S. J. Banks, K. T. Eddy, M. Angstadt, P. J. Nathan, and K. L. Phan, "Amygdala-frontal connectivity during emotion regulation," *Soc Cogn Affect Neurosci,* vol. 2, no. 4, pp. 303-12, Dec, 2007.

[30] Y.-Y. Lee, and S. Hsieh, "Classifying different emotional states by means of EEG-based functional connectivity patterns," *PloS one,* vol. 9, no. 4, pp. e95415, 2014.

[31] P. Y. Li, H. Liu, Y. Si, C. Li, F. Li, X. Zhu, X. Huang, Y. Zeng, D. Yao, Y. Zhang, and P. Xu, "EEG based emotion recognition by combining functional connectivity network and local activations," *IEEE Transactions on Biomedical Engineering*, 2019.

[32] P. Xu, X. Xiong, Q. Xue, P. Li, R. Zhang, Z. Wang, P. A. Valdes-Sosa, Y. Wang, and D. Yao, "Differentiating between psychogenic nonepileptic seizures and epilepsy based on common spatial pattern of weighted EEG resting networks," *IEEE Trans Biomed Eng,* vol. 61, no. 6, pp. 1747-55, Jun, 2014.

[33] M. Soleymani, J. Lichtenauer, T. Pun, and M. Pantic, "A Multimodal Database for Affect Recognition and Implicit Tagging," *IEEE Transactions on Affective Computing,* vol. 3, no. 1, pp. 42-55, 2012.

[34] S. Koelstra, C. Muhl, M. Soleymani, J. S. Lee, A. Yazdani, T. Ebrahimi, T. Pun, A. Nijholt, and I. Patras, "DEAP: A Database for Emotion Analysis Using Physiological Signals," *IEEE Transactions on Affective Computing,* vol. 3, no. 1, pp. 18-31, 2012.

[35] J. L. Lehmann, and F. Lynch, "Fully digital spectrum analyzer using time compression and discrete fourier transform techniques," Google Patents, 1975.

[36] Y. Peng, J. Y. Zhu, W. L. Zheng, and B. L. Lu, "EEG-based emotion recognition with manifold regularized extreme learning machine." pp. 974-977.

[37] W. L. Zheng, J. Y. Zhu, Y. Peng, and B. L. Lu, "EEG-based emotion classification using deep belief networks." pp. 1-6.

[38] V. Sakkalis, "Review of advanced techniques for the estimation of brain connectivity measured with EEG/MEG," *Computers in Biology & Medicine,* vol. 41, no. 12, pp. 1110-1117, 2011.

[39] J. P. Lachaux, E. Rodriguez, J. Martinerie, and F. J. Varela, "Measuring phase synchrony in brain signals," *Human brain mapping,* vol. 8, no. 4, pp. 194-208, 1999.

[40] S. Aydore, D. Pantazis, and R. M. Leahy, "A Note on the Phase Locking Value and its Properties," *Neuroimage,* vol. 74, 2013.

[41] S. C J, d. H. W, D. A, J. B F, M. I, v. C. v. W. A M, M. T, V. J P A, d. M. J C, and v. D. B W, "Graph theoretical analysis of magnetoencephalographic functional connectivity in Alzheimer's disease," *Brain,* vol. 132, no. Pt 1, pp. 213-224, 2009.

[42] B. Dai, and W. Yu, "Energy Efficiency of Downlink Transmission Strategies for Cloud Radio Access Networks," *IEEE Journal on Selected Areas in Communications,* vol. 34, no. 4, pp. 1037-1050, 2016.

[43] E. Van Diessen, T. Numan, E. Van Dellen, A. Van Der Kooi, M. Boersma, D. Hofman, R. Van Lutterveld, B. Van Dijk, E. Van Straaten, and A. Hillebrand, "Opportunities and methodological challenges in EEG and MEG resting state functional brain network research," *Clinical Neurophysiology,* vol. 126, no. 8, pp. 1468-1481, 2015.

[44] Z. J. Koles, M. S. Lazar, and S. Z. Zhou, "Spatial patterns underlying population differences in the background EEG," *Brain Topography,* vol. 2, no. 4, pp. 275, 1990.

[45] B. Blankertz, R. Tomioka, S. Lemm, M. Kawanabe, and K. R. Müller, "Optimizing spatial filters for robust EEG single-trial analysis," *IEEE Signal Processing Magazine,* vol. 25, no. 1, pp. 41-56, 2007.

[46] B. Blankertz, G. Dornhege, M. Krauledat, K.-R. Müller, and G. Curio, "The non-invasive Berlin brain–computer interface: fast acquisition of effective performance in untrained subjects," *NeuroImage,* vol. 37, no. 2, pp. 539-550, 2007.

[47] I. Syarif, A. Prugel-Bennett, and G. Wills, "SVM parameter optimization using grid search and genetic algorithm to improve classification performance," *Telkomnika,* vol. 14, no. 4, pp. 1502, 2016.

[48] J. J. Allen, and J. M. Coan, "Issues and assumptions on the road from raw signals to metrics of frontal EEG asymmetry in emotion," *Biological Psychology,* vol. 67, no. 1, pp. 183-218, 2004.

[49] L. Aftanas, N. Reva, A. Varlamov, S. Pavlov, and V. Makhnev, "Analysis of evoked EEG synchronization and desynchronization in conditions of emotional activation in humans: temporal and topographic characteristics," *Neuroscience and behavioral physiology,* vol. 34, no. 8, pp. 859-867, 2004.

[50] S. K. Sutton, and R. J. Davidson, "Prefrontal brain electrical asymmetry predicts the evaluation of affective stimuli," *Neuropsychologia,* vol. 38, no. 13, pp. 1723-1733, 2000.

[51] Y. Si, X. Wu, F. Li, L. Zhang, K. Duan, P. Li, L. Song, Y. Jiang, T. Zhang, Y. Zhang, and P. Xu, "Different Decision-Making Responses Occupy Different Brain Networks for Information Processing: A Study Based on EEG and TMS," *Cerebral Cortex.*

[52] E. Altenmüller, K. Schürmann, V. K. Lim, and D. Parlitz, "Hits to the left, flops to the right: different emotions during listening to music are reflected in cortical lateralisation patterns," *Neuropsychologia,* vol. 40, no. 13, pp. 2242-2256, 2002.

[53] H. Shahabi, and S. Moghimi, "Toward automatic detection of brain responses to emotional music through analysis of EEG effective connectivity," *Computers in Human Behavior,* pp. 231-239, 2016.

[54] R. Yuvaraj, M. Murugappan, U. R. Acharya, H. Adeli, N. M. Ibrahim, and E. Mesquita, "Brain functional connectivity patterns for emotional state classification in Parkinson's disease patients without dementia," *Behav Brain Res,* vol. 298, no. Pt B, pp. 248-60, Feb 1, 2016.

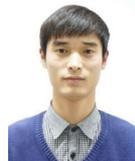

**Cunbo Li** is working for his Ph.D. degree in School of Life Science and Technology, University of Electronic Science and Technology of China. He graduated from Southwest University of Science and Technology, Mianyang, China, in 2017. His research mainly focuses on emotion classification, sleeping research, brain-computer interaction and deep learning.

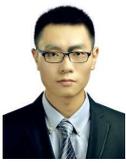
**Peiyang Li** received the Ph.D. degree in biomedical engineering from University of electronic science and technology of china, Chengdu, Sichuan, China, in 2018. He is now working in School of Bioinformatics, Chongqing University of Posts and Telecommunications, Chongqing, China. His research focuses on brain-computer interaction, convex optimization, machine learning and pattern recognition.

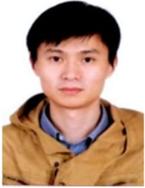
**Yangsong Zhang** received the Ph.D. degree in Signal and Information Processing from the School of Life Science and Technology, University of Electronic Science and Technology of China in 2013. He is currently an Associate Professor at the School of Computer Science and Technology, Southwest University of Science and Technology, China. Since 2016, he have been working as a Postdoctoral Research Fellow at University of Electronic Science and Technology of China (UESTC). His research interests include biomedical signal processing, brain–computer interface, machine learning, brain network analysis, etc.

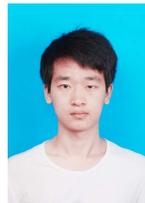
**Ning Li** is a graduate student currently in School of Life Science and Technology, University of Electronic Science and Technology of China. He graduated from School of Electrical Engineering, Zhengzhou University in 2019. His research interests include sleeping research, brain-computer interaction.

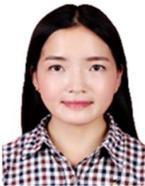
**Yajing Si** received the Master degree in psychology from the School of Psychology, Xin Xiang Medical University in 2016. She has been a doctor at School of Life Science and Technology, Center for Information in Medicine, University of Electronic Science and Technology of China (UESTC). Her research interests include decision-making cognition, attention, brain network analysis, etc.

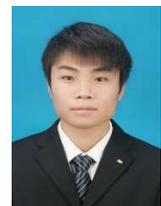
**Fali Li** received the bachelor degree in Biomedical Engineering from the School of Bioinformatics, Chongqing University of Posts and Telecommunications in 2013. And now he is working his Ph.D. degree in Biomedical Engineering from the School of Life Science and Technology, University of Electronic Science and Technology of China. His research interests include brain – computer interface, cognitive neuroscience, and brain network analysis, etc.

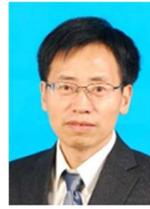
**Dezhong Yao** received the Ph.D. degree from the Chengdu University of Technology in 1991. He is the Fellow of the American Institute of Medical and Biological Engineering (AIMBE). He is currently the full professor at the School of Life Science and Technology, University of Electronic Science and Technology of China, China. His research interests include braininformatics, apparatus-Brain conversation, big neurodata mining, and neurocomputation for epilepsy.

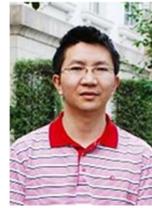
**Peng Xu** received the Ph.D. degree in Biomedical Engineering from the School of Life Science and Technology, University of Electronic Science and Technology of China in 2006. He is currently a Professor at the School of Life Science and Technology, University of Electronic Science and Technology of China, China. His research interests include the EEG inverse problem based on Lp Norm, brain–computer interface, machine learning, and brain network analysis, etc.